\documentclass[aps,pre,twocolumn,groupedaddress,amsmath,amssymb,showpacs]{revtex4-1}

\usepackage[pdftex]{graphicx}

\begin{document}

\title{Experimental measurements of stress redistribution 
in flowing emulsions}

\author{Kenneth W. Desmond and Eric R. Weeks}
\affiliation{Department of Physics, Emory University, Atlanta, Georgia 30322, USA}


\date{\today}

\begin{abstract}
We study how local rearrangements alter droplet stresses within
flowing dense quasi-two-dimensional emulsions at area fractions
$\phi \geq 0.88$.  Using microscopy, we measure droplet positions
while simultaneously using their deformed shape to measure droplet
stresses. We find that rearrangements alter nearby stresses in
a quadrupolar pattern: stresses on neighboring droplets tend to
either decrease or increase depending on location.  The stress
redistribution is more anisotropic with increasing $\phi$.
The spatial character of the stress redistribution influences
where subsequent rearrangements occur.  Our results provide direct
quantitative support for rheological theories of dense amorphous
materials that connect local rearrangements to changes in nearby
stress.
\end{abstract}


\maketitle

The flow of complex materials such as emulsions, foams, and
grains is the cumulative effect of many discrete particle
rearrangements \cite{Mason1996, Mason1999}. On shorter time and length
scales rearrangements result in local stress and velocity
fluctuations~\cite{Durian1995, Miller1996, Tewari1999, Dennin2004,
Lauridsen2004, Chen2012}, and the size and frequency of these
fluctuations are thought to relate to the macroscopic response
of the material~\cite{Picard2005, Kabla2007, Kabla2003}, but a
complete connection is yet unclear.

Simulations showed that particle rearrangements are initiated by a
build up of local inter-particle stresses that become unstable.
Particles rearrange and change the forces they exert on neighboring
particles, leading to an anisotropic redistribution of stress over
a few particle diameters~\cite{Kabla2003, Picard2004, Picard2005,
Kabla2007}. This redistribution is of significant interest because
it can increase the tendency for neighboring
particles to undergo a rearrangement \cite{Goyon2008, Picard2005,
Kabla2007}. This can lead to avalanches, where one rearrangement
alters the stress nearby and triggers a series of other
rearrangement~\cite{Debregeas2001, Pouliquen2009, Kamrin2012,
Kabla2007}.  Common models of the flow of complex materials, like
the Herschel-Bulkley equation, only consider the spatial gradient
in stress and strain rate between infinitesimal adjacent layers
of fluid sliding past each other \cite{Goyon2008, bocquet09, Becu2006}. However,
if a rearrangement can alter the stress farther away and trigger other
rearrangements, then the stress and strain rate couple to fluid
layers that extend far beyond the adjacent layers, requiring
a more complex treatment to properly model the flow. Some theories
assume forms for the stress redistribution to account
for the non-local effects, for example, the fluidity models for
emulsions \cite{Goyon2008, bocquet09} and grains
\cite{Pouliquen2009, Kamrin2012}.

To properly model the flow of these complex materials, the
nature of the stress redistribution must be known. Specifically,
(1) what is the magnitude of the stress redistribution, (2)
how far does it extend, and (3) how asymmetric is it? These
questions have been experimentally difficult to probe because of
the necessity to simultaneously image individual particles while
measuring the forces between them.  For this reason, most prior
experiments were limited to tracking particle positions without
knowledge of the stresses~\cite{Miller1996, Tewari1999, Dennin2004,
Lauridsen2004, chen10, besseling09, dijksman12}.  In this Letter,
we present experiments studying the flow of quasi-two-dimensional
emulsion droplets.  In our experiment we are able to simultaneously
observe both positions and inter-droplet forces~\cite{Chen2012,
Desmond2013}, allowing us to directly measure rearrangements and
stress redistribution.  We confirm that stress redistribution
occurs in a quadrupolar fashion close to what is predicted
\cite{Picard2004,Kabla2003}.

\begin{figure}[t]
\begin{center}
\includegraphics[width=3.4in]{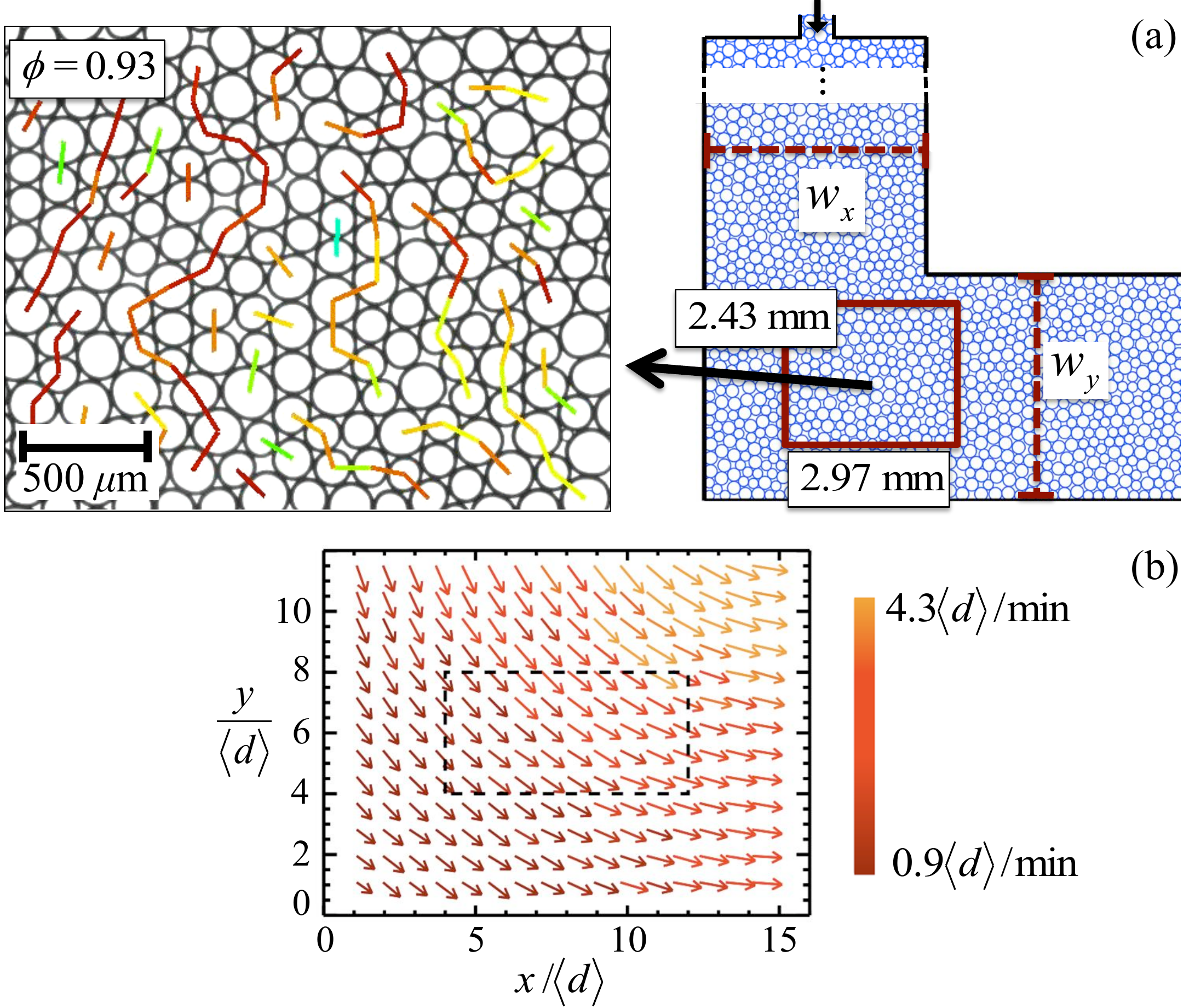}
\caption{(Color online). (a) Schematic of our two-dimensional
flow geometry. The
red rectangle indicates the region where the
flow is imaged with a snapshot of the flow of this region shown
to the left. On the snapshot, force chains \cite{Desmond2013} have been
drawn as lines, where the largest forces are in red.
(b) Average flow field of droplets in our field of view,
where $\langle{}d\rangle{} = 188$ $\mu$m is the mean droplet
diameter for these data. The length and color of each arrow
indicates the mean velocity. The dashed black
rectangle is the region in the field of view where T1 events are
considered for analysis. Both (a) and (b) correspond to the first row
of Table~\ref{table:Setup}.}
\label{fig:FlowGeometry}
\end{center}
\end{figure}

In our experiments, we confine bidisperse emulsions (mineral oil
in water) between two glass slides of dimensions 25 mm $\times$
75 mm separated by $\sim 100$ $\mu$m spacer (transparency
film).  We produce our droplets using a standard micro-fluidic
technique~\cite{Shan04} and stabilize them from coalescence with
Fairy soap \cite{Desmond2013}.  Their diameters are larger than the
sample chamber gap distance so that the system is quasi-2D.  A
schematic of our chamber is shown in Fig.~\ref{fig:FlowGeometry}(a)
which consists of two channels of widths $w_x \approx w_y$ that
meet at a right angle.  A syringe pump injects the emulsion into
the chamber, far upstream from the imaged region, as indicated in
Fig.~\ref{fig:FlowGeometry}.  In the corner the dropets change
direction, resulting in many rearrangements.  We take a total of 13
data sets at different area fractions $\phi$ ranging between 0.88 -
0.96 and at 4 different strain rates (see Table~\ref{table:Setup}
for experimental details).  We record images with a CCD camera at 2 images/s.
After the experiment, we post-process the images to identify and
track the droplets \cite{Desmond2013,crocker96}.

In Fig.~\ref{fig:FlowGeometry}, we show a typical flow field in
the region where the two channels meet, corresponding to where we
take all our data.  
Droplets flow faster near the upper right corner and are 
progressively slower with distance away from the corner. This
velocity gradient induces droplet rearrangements. The
gradient is non-uniform giving a strain rate that depends on space,
although locally it varies by no more than a factor of $\sim 2$
across the field of view for all our experiments.  The results we
will present do not vary with the local strain rate for the strain
rates we consider.  Accordingly, from the droplet trajectories
we determine the global (mean) strain rate within the region of our
observations and report this as $\dot{\gamma}_{global}$ for each
data set presented in Table~\ref{table:Setup}, similar to the
procedure in Ref.~\cite{chen10}.

\begin{table}[t]
\begin{center}
\begin{tabular}{cccccccc}
\# exp. & $\rho$ & $n_r$ & $R_s$ & $R_b$ & $w_x$ & $w_y$ & $\dot{\gamma}_{global}$ \\
 & & & [mm] & [mm] & [mm] & [mm] & $\left[\mbox{hr}^{-1}\right]$ \\
\hline
7 & 1.3 & 0.95 & 0.090 & 0.118 & 6.0 & 6.2 & 2.4 \\
1 & 1.3 & 0.95 & 0.090 & 0.118 & 6.0 & 6.2 & 1.3 \\
3 & 1.3 & 0.67 & 0.85 & 0.110 & 3.8 & 3.8 & 5.5 \\
2 & 1.3 & 0.67 & 0.85 & 0.110 & 3.8 & 3.8 & 2.7 \\
\hline
$\pm$  & 0.02 & 0.05 & 5\% & 5\% & 0.1 & 0.1 & 10\%
\end{tabular}
\end{center}
\caption{Attributes of the four setup/samples used.
The first column indicates the
number of experiments performed using that setup.  The size ratio
is $\rho{} = R_b/R_s$, where $R_b$ and $R_s$ 
are the average 2D radii of the big and small droplet species.
$n_r = N_b/N_s$ is the number ratio of the two species.  The last
row ($\pm$) indicates one standard deviation in the experimental
variations in each value.
}
\label{table:Setup}
\end{table}%

In addition to tracking the motion of droplets, we determine the
repulsive contact force $\vec{f}_{ij}$ between droplets $i$ and $j$
in contact.  This is done by relating the deformation of
each droplet's perimeter to the force using the method of Desmond
\textit{et al.}~\cite{Desmond2013}. Conceptually, the quasi-2D
droplets prefer to have circular outlines due to surface tension,
and their deformation away from circular relates to the forces
they feel.  We use our previously developed calibration technique
to determine the forces to 15\% uncertainty \cite{Desmond2013}.
In Fig.~\ref{fig:FlowGeometry}, we show some of the larger forces
between droplets.  We also determine the viscous droplet-droplet and
droplet-glass forces (see Ref.~\cite{Desmond2012} for details). The
largest viscous droplet-droplet forces are less than 1\% of the
mean repulsive force, and the largest viscous droplet-glass forces
are less than 4\% of the mean repulsive force. Since the viscous
forces are small, we ignore their contribution to the stress.

Using the repulsive contact forces, the stress tensor on each droplet at
any time can be computed using 
\begin{equation}
\overleftrightarrow{\sigma_i} = 1/A_i\displaystyle\sum_j \vec{r}_{ij}^{\mbox{ }c}\otimes{}\vec{f}_{ij}
\end{equation}
where $\vec{r}_{ij}^{\mbox{ }c}$ is a vector from the center
of droplet $i$ to the center of contact $ij$, $\otimes{}$
is an outer product, and $A_{i}$ is the Voronoi area around
droplet $i$~\cite{Cruz2005, Allen89}. 
The off-diagonal term of this tensor is the shear stress,
which depends on the orientation of the coordinate
system.  We rotate the coordinate system to maximize the time-
and droplet-averaged off-diagonal term of the stress tensor;
this rotation is $\approx 45^\circ$.  In this rotated coordinate
system, we use the instantaneous maximal off-diagonal element as
the instantaneous shear stress $\sigma_i(t)$ on each droplet.

\begin{figure}[t]
\begin{center}
\includegraphics[width=3.4in]{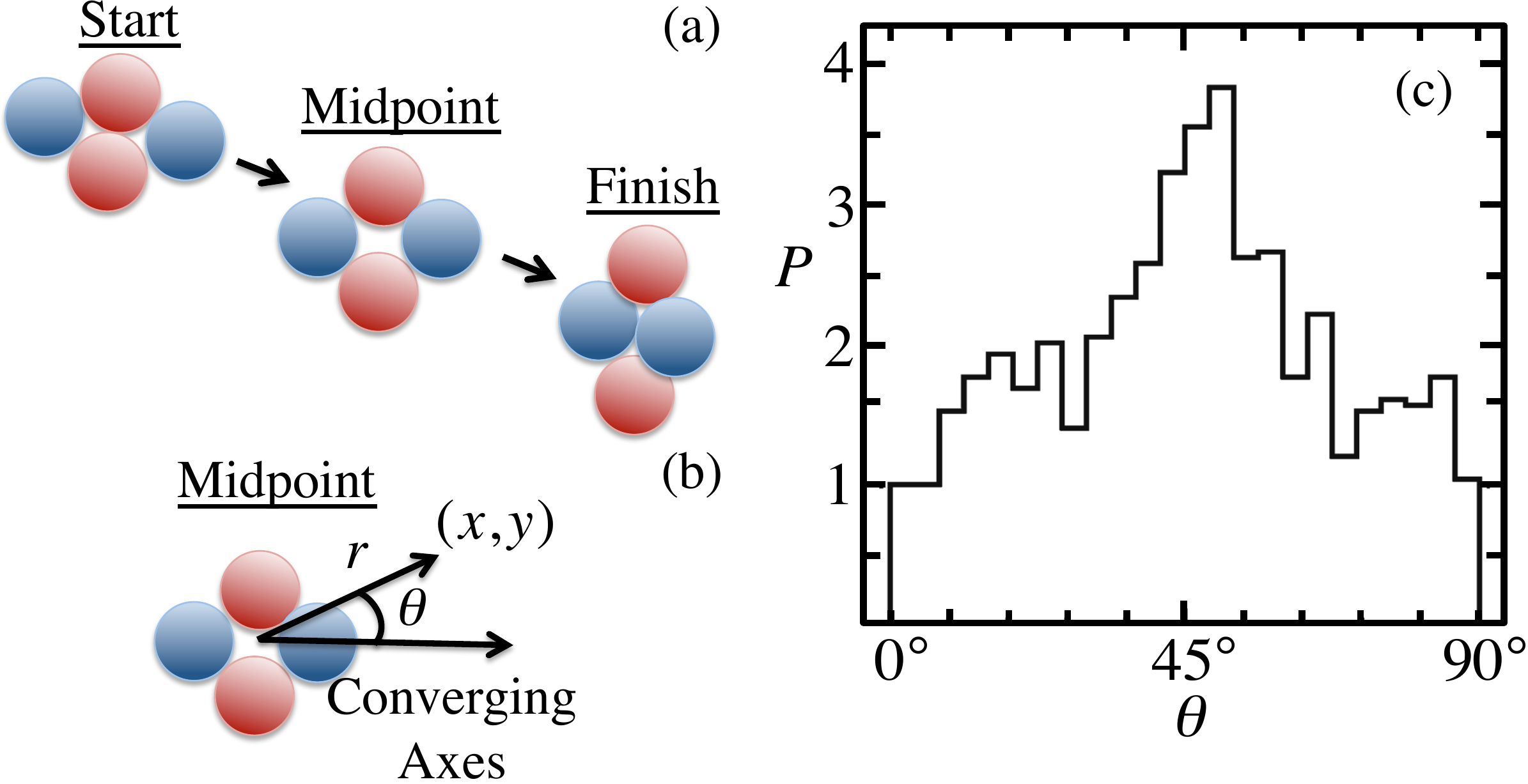}
\caption{(Color online). (a) T1 event, where 4 droplets exchange
being neighbors. (b) Coordinate system relative to midpoint of T1
event. The converging axis is the line joining the centers of the
two droplets moving closer together. The direction of the axis is
arbitrary. Relative to the axis we define the Cartesian coordinates
$(x, y)$ and the polar coordinates $(r,\theta)$. (c) Probability to
find two successive T1 events located at an angle $\theta$ relative
to the converging direction of the first particle rearrangement for all
our data at any $\phi$.}
\label{fig:OrientationBtwEvents}
\end{center}
\end{figure}

In 2D systems, the simplest topological rearrangements involve
neighbor exchanges of four droplets~\cite{Lundberg08},
known as a T1 event.  This exchange is shown in
Fig.~\ref{fig:OrientationBtwEvents}(a): two neighboring droplets
become next-nearest neighbors while two next-nearest neighbors
become neighbors.  To uniquely define the start
and end of a T1 event, we only consider T1 events that lead to
a stress relaxation on the four rearranging droplets, and define
the start as the point in time where the stress is maximal before
the neighbor exchange and the end as the minimum in stress after
the neighbor exchange. Between the start and the end of the T1
event the stress decreases by $\delta{}\sigma_{\mathrm{T1}}$,
which not surprisingly varies between rearrangements.

\begin{figure}[t]
\begin{center}
\includegraphics[width=3.4in]{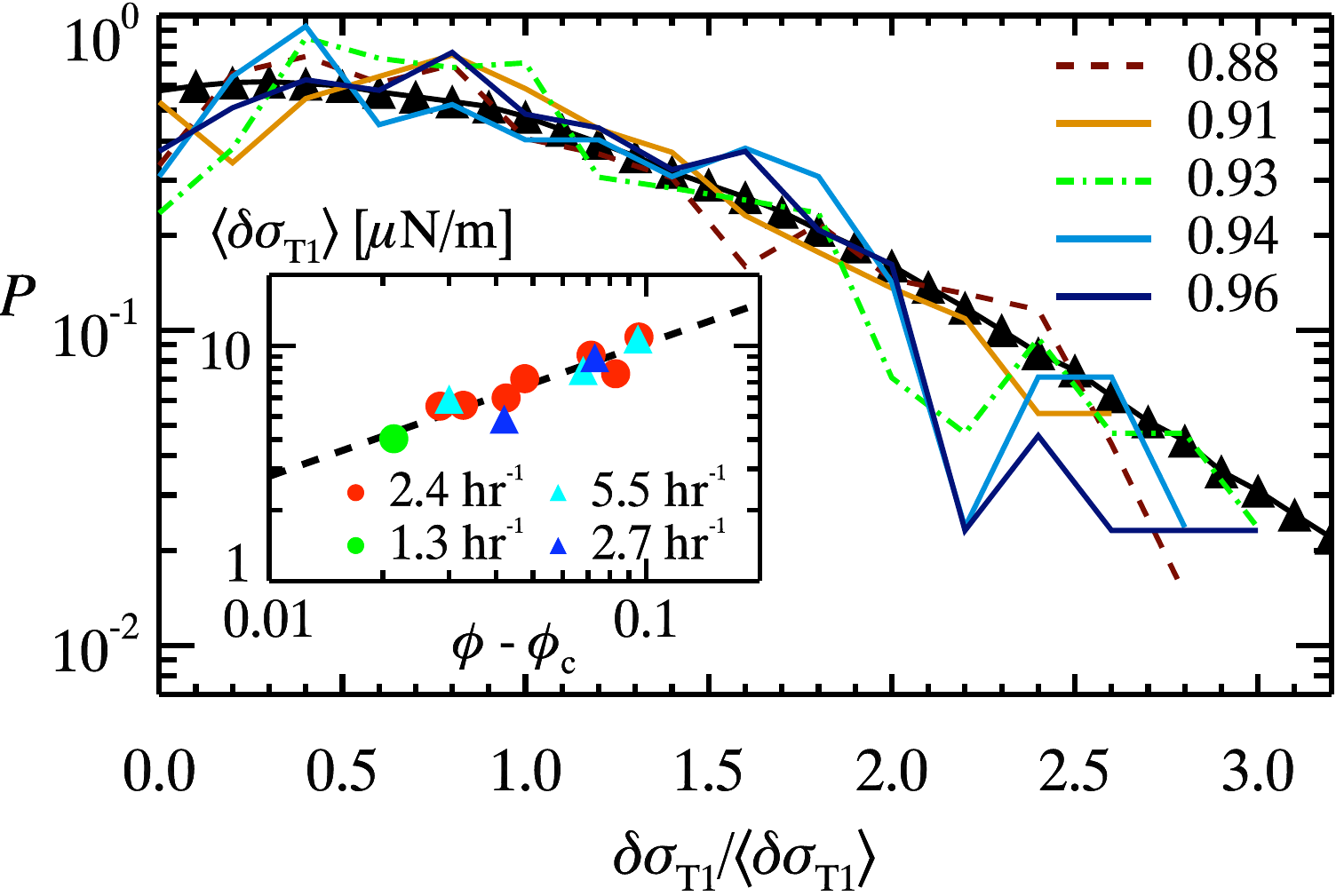}
\caption{(Color online). Distribution of stress decrease on
the rearranging droplets. The different colored curves indicate
different area fractions. The solid curves are data at $\dot{\gamma}
= $ 5.5 hr$^{-1}$, the dashed lines at $\dot{\gamma} = $ 2.7
hr$^{-1}$, and the dashed dot lines at $\dot{\gamma} = $ 2.4
hr$^{-1}$. The filled triangle black curve is the distribution
of individual droplet shear stresses $\sigma_i$ in a sample at $\phi =
0.93$, normalized by $\langle \sigma_i \rangle$. 
Inset: Average stress relaxation with area fraction. The
legend indicates data at different strain rates. The black dashed
line is a power law fit $\langle{}\delta{}\sigma{}\rangle{} = A(\phi
- \phi_c)^\beta$, where $\beta = 0.57$ and $A = $ 38 $\mu$N/m. }
\label{fig:StressDrops}
\end{center}
\end{figure}

To examine the variability of $\delta\sigma_{\mathrm{T1}}$, we show
the probability distributions of $\delta\sigma_{\mathrm{T1}}$
as the curves in Fig.~\ref{fig:StressDrops}.  These
distributions have a Gaussian-like shape and peak near
$0.5\langle{}\delta\sigma_{\mathrm{T1}}\rangle$.
After normalizing by the mean stress drop
$\langle{}\delta\sigma_{\mathrm{T1}}\rangle$, all distributions
overlap well regardless of $\phi$ and $\dot{\gamma}$. This
suggests that in the quasistatic limit, fluctuations in
local stress due to structural relaxation follow a universal
distribution, although we cannot rule out subtle differences that
might be below our resolution.  Intriguingly, the distributions
$P(\delta\sigma_{\mathrm{T1}})$ look similar to the distribution of
the instantaneous individual droplet stresses $P(\sigma_i$), plotted
as triangles in Fig.~\ref{fig:StressDrops} (at a representative
area fraction $\phi=0.93$; the shape does not vary significantly
at different $\phi$).  This seems sensible, suggesting that the
size of the stress drops correlates with the size of the stresses
present in the system.

While the variation in $\delta\sigma$ about the mean is
independent of the area fraction, the mean stress drop is
not. In the inset of Fig.~\ref{fig:StressDrops}, we show that
$\langle{}\delta\sigma\rangle$ increases with area fraction relative
to the jamming point $\phi_c = 0.86$, where $\phi_c$ was measured
in Ref.~\cite{Desmond2013}.  The data for different strain rates
overlap, as we are in the quasi-static regime.  We fit the data
using $\langle{}\delta\sigma\rangle \sim (\phi - \phi_c)^\beta$
with $\beta = 0.57$, similar to scaling behavior found for the
shear modulus, pressure, and coordination number~\cite{Durian1995,
Durian1997}.

Thus far we focused on the stress drop averaged over the four
droplets defining the T1 event.  We next examine how the T1
event redistributes stress on the rest of the sample.  To do
this, we define a coordinate system for each T1 event, as shown
in Fig.~\ref{fig:OrientationBtwEvents}(b), using the converging
direction of droplets as the $x$-axis and the diverging direction
as the $y$-axis.  Here $(x, y)$ or $(r, \theta)$ measure locations
relative to the center of the T1 event.

The stress redistribution around a T1 event can be characterized by
a stress propagator $\Pi(x, y)$, defined as $\delta{}\sigma(x,y) =
\Pi(x, y)\delta\sigma_{\mathrm{T1}}$, where $\delta{}\sigma(x,y)$
is the change in stress on a droplet located a distance $(x,y)$ away
from the center of the T1 event that has a particular stress drop
of size $\delta\sigma_{\mathrm{T1}}$. Given a stress fluctuation on the
rearranging droplets, $\Pi(x, y)$ quantifies the mean size of the stress
fluctuation at $(x,y)$. Recall, $\delta{}\sigma_{\mathrm{T1}}$
measures the amount the stress has decreased, and thus, when $\Pi(x, y)$
is positive the stress at $(x, y)$ decreased. Here we will 
focus on the stress propagator $\Pi(x,y)$ averaged over 200-400 T1
events (we ignore the angular brackets in our notation).

\begin{figure}[t]
\begin{center}
\includegraphics[width=3.4in]{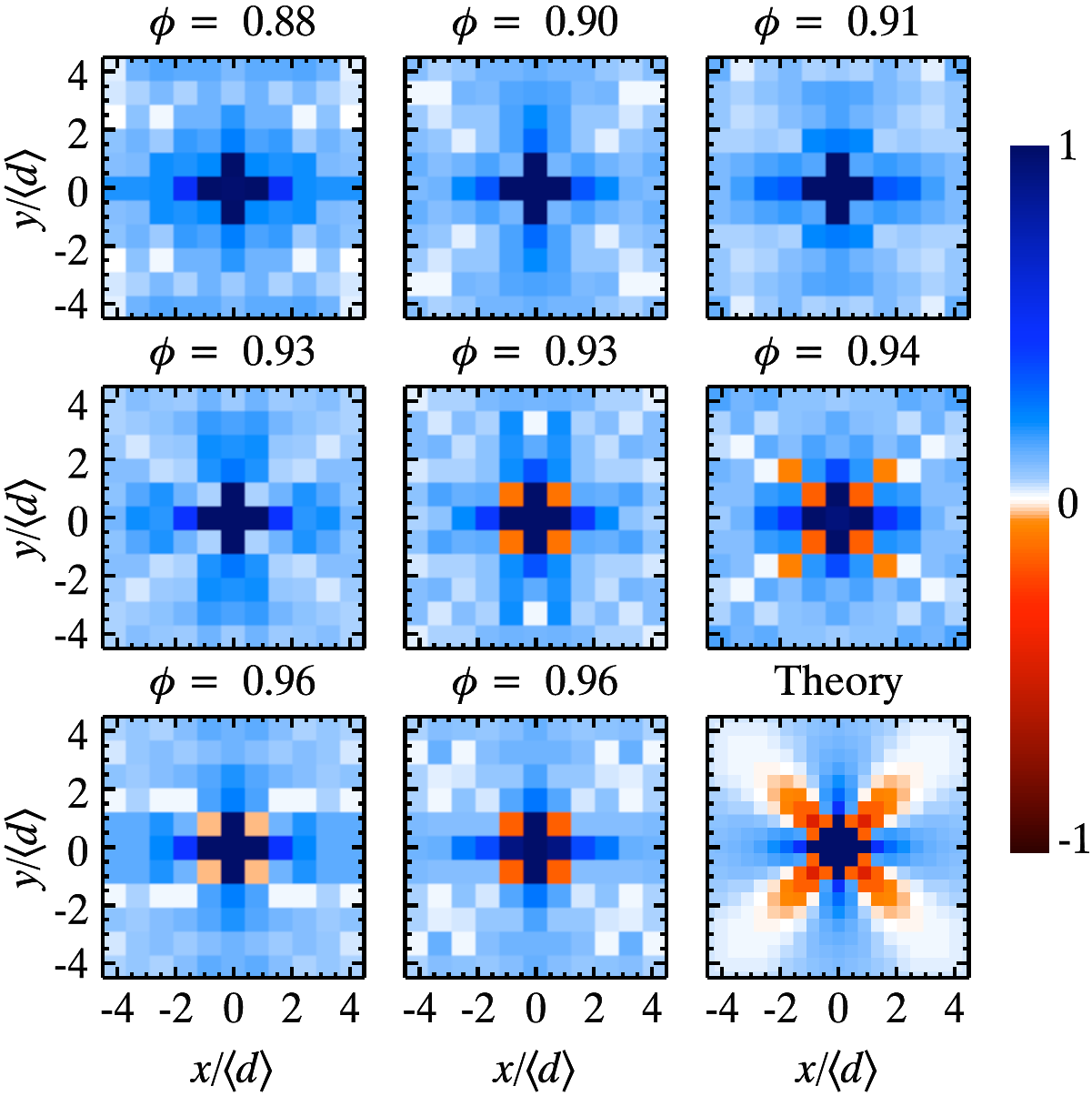}
\caption{(Color online). Average stress propagator for 8
different samples. The last panel is the theoretical
stress propagator \cite{Picard2004}. Blue indicates an average
stress decrease on droplets located a distance $x$ and $y$ away
from the center of the particle rearrangement, and red indicates a stress
increase. The relative position to the center of the particle rearrangement is measured in units of average droplet diameters. Both $\phi = 0.93$ panels and both $\phi = 0.96$ panels are two different samples with the same $\phi$, indicating that the stress propagator is not completely reproducible between experiments.}
\label{fig:StressPropagator}
\end{center}
\end{figure}

In Fig.~\ref{fig:StressPropagator}, we show the measured stress
propagator for a range of area fractions.  To improve statistics,
we impose four fold symmetry on $\Pi$.  (Prior to imposing symmetry,
the raw data have four fold symmetry within the statistical noise
limit.) The data in Fig.~\ref{fig:StressPropagator} show that the
stress can decrease (positive $\Pi$, blue) or increase (negative
$\Pi$, orange) on nearby droplets depending on their relative
position to the T1 event. Along the converging and diverging axes
we find that for all $\phi$ the stress decreases on droplets near
the rearrangement, although along the directions $45^{\circ{}}$
to the converging and diverging directions, the stress change
depends on $\phi$. At low area fractions, the stress propagators
show that a rearrangement tends to decrease the stress on all
nearby droplets, while at larger area fractions, droplets along
the diagonal directions tend to have their stress increased.

Picard \textit{et al.}~\cite{Picard2004} model a T1 event as
a localized region of size set by the mean droplet diameter
$\langle{}d\rangle{}$ undergoing pure shear within a continuous
elastic material. They predict a quadrupolar field for the stress
propagator obeying $\Pi = (\langle d \rangle/r)^2\cos(4\theta)$,
shown in the bottom right panel of Fig.~\ref{fig:StressPropagator}.
Along the converging and diverging directions, the stress on
neighboring droplets decreases, while the stress on the nearby
droplets at $\sim$45$^\circ{}$ to the converging direction
increases. In simulations on dry foams ($\phi = 1$), Kabla
\textit{et al.}~\cite{Kabla2003} observed a qualitatively similar
quadrupolar field in the stress propagator.

\begin{figure}[t]
\begin{center}
\includegraphics[width=3.4in]{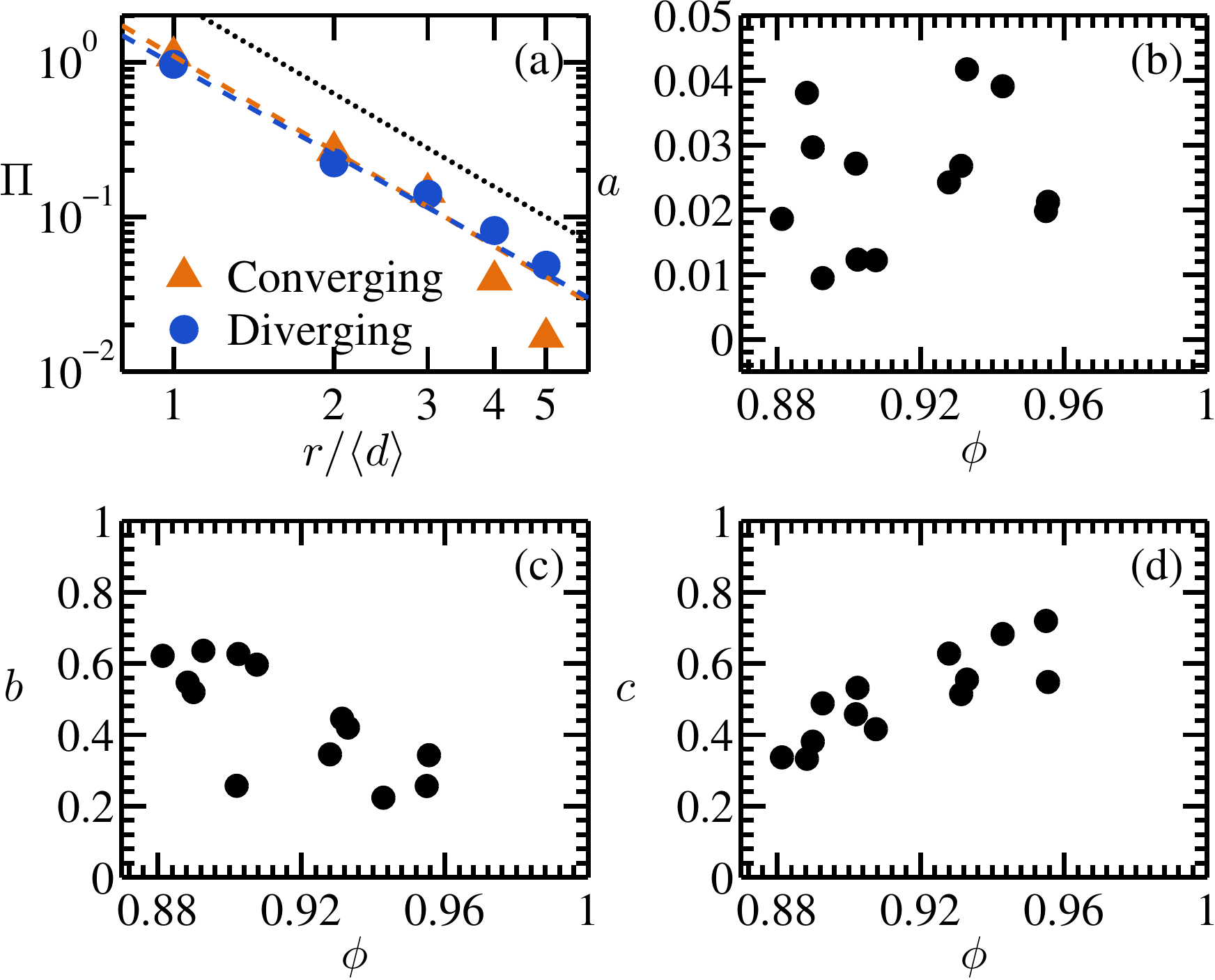}
\caption{(Color online). 
(a) Decay of stress propagator along the converging and diverging
directions in a sample at $\phi = 0.89$, The colored dashed lines
are power law fits of the form $C/r^\alpha$, where $C = $ 1.09
and 0.97, and $\alpha = $2.04 and 1.94 for the converging and
diverging directions respectively. The black dashed line is a
power law with exponent -2 as a guide to the eye.
(b)-(d) The stress propagator for each data set
is fit to $\Pi_{model} = a + b(\langle{}d\rangle/r)^2 +
c(\langle{}d\rangle/r)^{2}\cos(4\theta)$, and the fitting
parameters $a$, $b$, and $c$ are shown as functions of area fraction.
}
\label{fig:StressPropagatorFits}
\end{center}
\end{figure}

We first test Picard \textit{et al.}'s prediction of a
$1/r^2$ scaling by looking at the decay in $\Pi$ along the
converging and diverging directions.  An example is given
in Fig.~\ref{fig:StressPropagatorFits}(a).  The data are
consistent with $1/r^2$ decay, indicated by the dotted line.
Next we look at the anisotropy of the stress propagator.
Qualitatively, we find that as the area fraction approaches
unity, the experimentally measured stress propagator appears to
approach a quadrupolar field. However, at lower area fractions
the stress propagator appears more isotropic.  To quantify
this, we add onto the prediction of Picard \textit{et al.}
two isotropic terms, and fit the raw data (without imposing
4-fold symmetry) to $\Pi_{model} = a + b(\langle{}d\rangle/r)^2 +
c(\langle{}d\rangle/r)^{2}\cos(4\theta)$, with fit parameters $a$,
$b$, and $c$. This is an empirical fit that works well for all our
data; we do not rule out the possibility of other forms for $\Pi$.

The fitting parameters are shown in
Fig.~\ref{fig:StressPropagatorFits}(b) - (d).  The monopole term $a$
is slightly positive with no trend in $\phi$.  $a>0$ shows that the
global (space-averaged) stress tends to decrease during a T1 event,
supporting the idea that rearrangements are a mechanism to relax
the global stress.  Qualitatively, this is seen in the overall
``blue" colors of the panels in Fig.~\ref{fig:StressPropagator}.
Examining the other fitting parameters, we find that $b$ (the
isotropic term) decreases with $\phi$, while $c$ (the quadrupolar
anisotropic term) increases with $\phi$. These trends indicate a
transition from an anisotropic stress relaxation to an isotropic
one as the jamming area fraction ($\phi_c=0.86$) is approached
from above. Simulations have shown \cite{Ellenbroek2006}, that
with increasing $\phi$ above $\phi_c$, the response of grains to
a localized force is more affine (elastic-like). It is reasonable
that the prediction of Picard {\it et al.} works well for samples
in the limit $\phi \rightarrow 1$, since these cases are closer
to their assumption of a continuous elastic material.

The above analysis shows that along the converging and diverging
axes the stress drops significantly, while droplets located $\sim
45^\circ$ from the converging axis show either a slight decrease
or a large increase in stress.  One would expect subsequent
rearrangements to be more likely where the stress has increased,
and this is what we find.  We measure the location of the next
T1 event in the same coordinate system used to compute $\Pi$. In
Fig.~\ref{fig:OrientationBtwEvents}(c), we plot the probability
to find the next event at angle $\theta$ relative to the previous.
($P(\theta)$ overlaps well for all the data sets, so to
improve the statistics we use the average of all the data.)
The peak near $45^\circ$ confirms that events are biased to occur
along the directions where the stress typically increases during
a T1 event.

Our experiments demonstrate that simple rearrangement events
(T1 events) relax the stress on neighboring droplets.  At higher
area fractions, we observe that this stress relaxation has
a quadrupolar character, confirming theoretical predictions
\cite{Picard2004}.  One implication is that T1 events increase
the stress felt by nearby droplets along certain directions, and
subsequent T1 events are more likely to occur in those regions
with increased stress.  Our observations provide direct evidence
that on a droplet scale, flow is spatially heterogeneous with a
connection between rearrangements in one location and stresses felt
in other locations.  While the spatially heterogeneous nature of
rearrangements in amorphous materials has been well-established by
prior work \cite{Miller1996, Tewari1999, Dennin2004, Lauridsen2004,
chen10, besseling09, dijksman12}, this is the first quantitative
experimental link connecting such rearrangements to the internal
stresses, as expected by a variety of theories of rheology of
complex fluids \cite{Picard2005, Kabla2007, Kabla2003, Picard2004,
Goyon2008, Debregeas2001, Pouliquen2009, Kamrin2012, bocquet09}.

This work was supported by the National Science Foundation (grant
CBET-0853837).


%

\end{document}